\documentclass[showpacs,preprintnumbers,amssymb,twocolumn]{revtex4}        
\usepackage{graphicx}
\usepackage{color}
\usepackage{dcolumn}
\usepackage{bm}
\usepackage{longtable}
\usepackage{graphics}
\date{\today}
\begin{document}
\title{Information transfer using a single particle path-spin hybrid entangled state}
\author{T. Pramanik\footnote{tanu.pram99@bose.res.in}$^1$, S. Adhikari\footnote{satyabrata@bose.res.in}$^1$, A. S. Majumdar\footnote{archan@bose.res.in}$^1$,  Dipankar Home\footnote{dhome@bosemain.boseinst.ac.in}$^2$ and Alok Kumar Pan\footnote{apan@bosemain.boseinst.ac.in}$^2$}

\address{$^{1}$ S. N. Bose National Centre for Basic Sciences,
Salt Lake, Kolkata 700 098, India}

\address{$^{2}$ CAPSS, Department of Physics, Bose Institute, Sector-V, 
Salt Lake, Kolkata
700 091, India}
\begin{abstract}
The path-spin entangled state of a single spin-$1/2$ particle is considered
which is generated by using a beam-spitter and a spin-flipper. Using this 
hybrid entanglement at the level of a single particle
as a resource, we formulate a protocol for transferring of the state of 
an  unknown qubit to a distant location. Our scheme is implemented
by a sequence of unitary operations along with suitable spin-measurements, 
as well as by using classical
communication between the two spatially separated parties. This protocol, 
thus, demonstrates the
possibility of using intraparticle entanglement as a physical resource
for performing information theoretic tasks.
\end{abstract}
\maketitle

Quantum entanglement is a crucial ingredient in
the storage and distribution of quantum information in the currently 
vibrant research area of quantum information\cite{peres1}. Historically,
the first fundamental implication of entanglement was noticed
in terms of position-momentum  variables
\cite{EPR}, and was later extended for the discrete spin variables
\cite{bohm}. In recent times, the theory of entanglement has been much
studied for systems described by
Hilbert spaces for the 
discrete variables\cite{review1} on the one hand,
and for those corresponding to continuous variables
\cite{review2} on the other. Several
interesting information processing protocols
such as quantum teleportation \cite{bennett1}, dense-coding \cite{denscod}
and cryptography \cite{crypto} have been developed for spin entangled states,
as well as for position-momentum entangled states \cite{teleport,cvcrypt}.

Further, it needs to be noted that in the larger context of investigating
various ramifications of quantum entanglement and its applications to a 
wide range of diverse phenomena such as phase
transitions in condensed matter systems \cite{condmat} and black hole
physics \cite{bhinf}, the study of hybrid entanglement between
the dynamical variables belonging to different Hilbert spaces
such as those corresponding to path (or linear momentum) variables 
on one hand, and spin variables on the other, is
particularly relevant. However, surprisingly, even though quantum
mechanics allows for the existence of hybrid entangled states connecting
Hilbert spaces with distinctly different properties, the possibility of
physical realization of such states is only
beginning to be appreciated \cite{hybrid,polmom,polangmom,blasone}.

Even more interesting is the idea of generation of intra-particle entanglement
between different degrees of freedom of the {\it same particle}.
The entanglement between the polarization and the linear momentum
of a single photon \cite{polmom}, and also the polarization and the angular
momentum of a single photon \cite{polangmom} has been demonstrated
experimentally. Further, it has been shown recently how flavor oscillations
of neutrinos could be related to multimode entanglement of single particle
states \cite{blasone}. The idea of creating entanglement between the 
path and the
spin degrees of freedom for a single spin-$1/2$ particle was proposed
earlier in order to demonstrate contextuality in quantum mechanics\cite{home}. Such path-spin hybrid entangled states for single
neutrons have also been realized experimentally \cite{hasegawa}.
Recently, it has been shown how intra-particle hybrid entanglement could be
swapped onto the standard inter-particle entanglement of two 
qubits\cite{adhikari}.

Since intra-particle entanglement between different degrees of freedom
is confined locally with a single particle, it should be easier to preserve,
at least in principle, against dissipative effects. It is
then natural to ask the question as to whether such hybrid entanglement
between different degrees of freedom of the same particle could be used as
resource for information processing. At the outset such an idea seems
difficult to implement,  since this entanglement is not delocalised
between two regions in a way that is amenable for exploiting as a resource.
It may be relevant to recall here the interesting debate in the literature 
regarding demonstration of nonlocality at the level of an entangled state of a 
single particle \cite{vanenk}.
In this Letter we devise a protocol for using hybrid entanglement at the level
of a single particle for quantum information processing. 
We show how the path-spin entanglement of a single spin-$1/2$
particle could be used as a resource for transferring the state of
an unknown qubit at a distant location. In order to realize such a scheme, 
we first discuss a method
to set-up the intra-particle hybrid path-spin entanglement using a
beam-splitter and a spin-flipper. Our protocol for information transfer
then proceeds with a series of operations performed by the two distant
parties (Alice and Bob) including unitary transformations, appropriate
measurements using Stern-Gerlach devices, and classical communications.
Our scheme is pictorially illustrated in Fig. 1.
\begin{figure}[h]
{\rotatebox{270}{\resizebox{6.0cm}{8.0cm}{\includegraphics{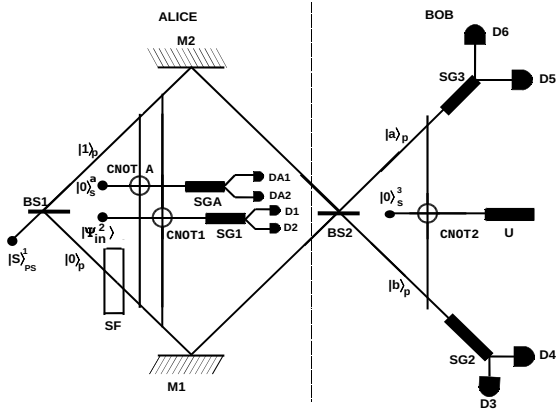}}}}
\caption{\footnotesize A spin-$1/2$ particle 
(labelled as particle 1) with an 
initial spin polarised state $|\uparrow\rangle_z$ falls on the beam-splitter 
BS1. A spin-flipper is placed along the reflected channel 
(labelled by $|0\rangle_p$). A CNOT operation is
performed by Alice involving this particle and another particle 
(labelled as ``a''. A
second CNOT operation is performed involving particle 1 and her second 
particle (2)teleportnew
which is in an unknown state $|\psi^{in}\rangle$. Alice
sends the particle (1) to Bob who lets this particle fall on the
beam-splitter BS2.
A CNOT operation is performed by Bob involving this particle and another
 particle (3) possessed by him (labelled as  $|0\rangle_s^3$).
Bob measures the spin of the particle (1) using the Stern-Gerlach
devices SG2 and SG3.  Alice then measures
the spins  of the
particle (2) and particle (a) using the Stern-Gerlach devices SG1 and SGA
along the $z$-axis.
According to the results of the spin measurements through SG1 and SGA 
(communicated 
classically by Alice to Bob), and through SG2 or SG3 (as performed by Bob),
he applies a  suitable unitary operation ( U ) on his particle (3) in
order to recreate the original state $|\psi_{in}\rangle$ possessed by Alice.}
\end{figure}

Let us consider an ensemble of spin-$1/2$ particles,
all corresponding to an initial
spin polarized state along the $+\widehat {z}-axis$ (denoted
by $\left|\uparrow_{z}\right\rangle$). Taking into consideration the path (or
position) variables of the particles, one can write down the joint path-spin
state for ensemble as
\begin{eqnarray}
|S\rangle^1_{ps} = |\psi_{0}\rangle_{p} \otimes |\uparrow_{z}\rangle_{s}
\label{redstate}
\end{eqnarray}
where the subscripts $p$ and $s$ refer to the path and spin variables,
respectively.
The particles are then allowed by Alice to fall on a  beam-splitter (BS1). 
Since the beam-splitter acts only on the path-state without
affecting the spin-state of the particles, the input-output relation is
described by
the unitary transformation 
\begin{eqnarray}
\left(\begin{matrix}{|\psi_{0}\rangle_{p}   \cr
|\psi_{0}^{\perp}\rangle_{p}}\end{matrix}\right) =
\left(\begin{matrix}{\alpha & i\beta \cr -\beta &
i\alpha}\end{matrix}\right)
\left(\begin{matrix}{|1\rangle_{p} \cr
|0\rangle_{p}}\end{matrix}\right) \label{beamsplit}
\end{eqnarray}
with $\alpha^2 + \beta^2 =1$ ($\alpha$ and $\beta$ are real),  
where  $\alpha^{2}$ and $\beta^{2}$ are the
reflection and transmission probabilities 
respectively, and the
reflected  or the transmitted channel are designated
by $|0\rangle_p$ or $|1\rangle_p$ respectively.  
In the following argument a crucial role is provided by the mutually
orthogonal path states $|0\rangle_p$ and
$|1\rangle_p$ which are  eigenstates of the projection
operators $P(|0\rangle_p)$ and $P(|1\rangle_p)$ respectively. These projection
operators  can be regarded as corresponding to the observables that pertain
to the determination of \textit{`which channel'} a particle is found to be in.
For example, the results of such  a measurement for the reflected (transmitted)
channel with binary alternatives are given by the eigenvalues of
$P(|0\rangle_p)$ ($P(|1\rangle_p)$); the eigenvalue $+1 (0)$ corresponds to a
neutron being found (not found) in the channel represented by
$|0\rangle_p$ ($|1\rangle_p$).

The state of a particle emergent from BS1 can be written as
\begin{eqnarray}
|S\rangle^1_{ps}\rightarrow |S_{1}\rangle^1_{ps} =
(\alpha|1\rangle_{p}+i\beta|0\rangle_{p}) \otimes
|\uparrow_{z}\rangle_{s} \label{transformedstate}
\end{eqnarray}
We can identify the state vectors
$|0\rangle_{p}$, $|1\rangle_{p}$, $|\uparrow_{z}\rangle_{s}$,
and $|\downarrow_{z}\rangle_{s}$ as
\begin{eqnarray}
|0\rangle_{p}\equiv\left(\begin{matrix}{1 \cr 0
}\end{matrix}\right), \>\>
|1\rangle_{p}\equiv\left(\begin{matrix}{0 \cr 1
}\end{matrix}\right),\nonumber\\
|\uparrow_{z}\rangle_{s}\equiv|0\rangle_{s}=\left(\begin{matrix}{1
\cr
0}\end{matrix}\right), \>\>
|\downarrow_{z}\rangle_{s}\equiv|1\rangle_{s}=\left(\begin{matrix}{0
\cr 1 }\end{matrix}\right)\label{notation}
\end{eqnarray}
Note that though we are considering here a single particle, the dichotomic
path and spin variables enable it to be viewed effectively as two qubits.
Next, suppose the particles in the channel corresponding to
$|0\rangle_p$ pass
through a spin-flipper (SF) (that contains a uniform magnetic field along,
say, the $+\widehat x$-axis) which flips the state
$\left|\uparrow\right\rangle_{z}$ to $\left|\downarrow\right\rangle_{z}$.
The resultant path-spin entangled state (using the above notation)
can be written as
\begin{eqnarray}
|S_{2}\rangle^1_{ps}= U_{CNOT} |S_{1}>_{ps} = \left(\begin{matrix}{0  \cr i\beta 
\cr \alpha \cr
0}\end{matrix}\right){}
\label{CNOT}
\end{eqnarray}
Or, in other words, as a consequence of introducing a spin-flipper
in one of the channels, Alice
now possesses the path-spin entangled state given by
\begin{eqnarray}
|S_{2}\rangle^1_{ps} &=& \alpha|\uparrow_{z}\rangle_{s}\otimes|1\rangle_{p}
+i\beta|\downarrow_{z}\rangle_{s}\otimes|0\rangle_{p}\nonumber\\
&\equiv&
\alpha|0\rangle_{s}\otimes|1\rangle_{p}+i\beta|1\rangle_{s}\otimes|0\rangle_{p}
\label{pathspinent}
\end{eqnarray}
Note that the above path-spin entanglement between the spin variables and
the path observables of a spin-$1/2$ particle -  the `intraparticle
 entanglement' involved here is distinct from the usually discussed
`interparticle entanglement', say,  between the spin variables of two
spatially separated particles, or even the recently discussed \cite{hybrid}
`hybrid entanglement' between the polarization of one photon and the
spin of another spatially separated photon. Further, the entangled state
generated here is also different
from several single particle entangled states involving the entanglement
of similar modes (e.g., Fock states \cite{vanenk}, or neutrino flavor
states \cite{blasone}) discussed earlier in the literature. The present
state corresponds to a single particle hybrid entangled state where the
entanglement is between the spin and path variables of the same particle. 

Our goal is to now use this entangled state as a resource for performing
the information transfer of the state of another particle (2) held by 
Alice. The 
state of particle 2 is given by
 \begin{eqnarray}
|\psi^{in}\rangle^2 = \gamma|0\rangle^2_s + \delta|1\rangle^2_s
\label{state}
\end{eqnarray}
where $\gamma^2 + \delta^2 = 1$.
At this stage we introduce another auxiliary particle possesed by Alice
prepared in the state $|0\rangle_s^a$ .
The role of this particle is to ensure, as will be clear later,  that the
information about the state to
be transferred is not lost even if the particle to be sent to Bob is lost
in transit. 
The state transfer protocol begins with Alice making a CNOT operation, where 
the first particle's spin state is the control qubit and the auxiliary 
particle's 
spin state is the target qubit. After the CNOT operation the combined state 
of the first and the auxiliary particle is given by
\begin{eqnarray}
|S_3\rangle_{pss}^{1a}=(\alpha||1\rangle_p^1|0\rangle_s^1|0\rangle_s^a+i\beta||0\rangle_p^1|1\rangle_s^1|1\rangle_s^a)
\label{auxstate}
\end{eqnarray}
Next, Alice makes another CNOT operation where 
the first particle's spin state is the control qubit and the second
particle's 
spin state (\ref{state}) is the target qubit.
The resultant combined state
of the three particles (first, second, and auxiliary) is now given by  
\begin{eqnarray}
|\psi\rangle^{12a}&=&
\alpha\gamma|1\rangle_p^1|0\rangle_s^1|0\rangle_s^2|0\rangle_s^a + i\beta\gamma|0\rangle_p^1|1\rangle_s^1|1\rangle_s^2|1\rangle_s^a \nonumber \\
&+&\alpha\delta|1\rangle_p^1|0\rangle_s^1|1\rangle_s^2|0\rangle_s^a + i\beta\delta|0\rangle_p^1|1\rangle_s^1|0\rangle_s^2|1\rangle_s^a
\label{redstate}
\end{eqnarray}
Note here that the CNOT operations are performed by Alice on whichever path
(transmitted or reflected) that is taken by her particle 1 on emerging from
the beam splitter. This has to be done in order to use the path-spin entangled
state as the teleportation channel. Performing any such operation before 
particle 1 falls on the beam splitter is of no use for teleportation, since
the path-spin entanglement itself is created after the particle emerges from
the beam splitter.

Alice then sends her first particle (in a path-spin entangled state) to Bob. 
After Bob confirms  that he has received the particle,
Alice measures the spin of her second particle along the $z$-axis using
her Stern-Gerlach device (SG1 in FIG. 1), and also measures the spin of her
auxiliary particle also along the $z$-axis using her Stern-Gerlach device
(SGA in FIG. 1).
Eq.(\ref{redstate}) can be rewritten as
\begin{eqnarray}
|\psi\rangle^{12a}&=&
\frac{1}{\sqrt{2}}[(\alpha\gamma|1\rangle_p^1|0\rangle_s^1 +i\beta\delta|0\rangle_p^1|1\rangle_s^1)|0\rangle_s^2|0_x\rangle_s^a\\ 
&+& (\alpha\gamma|1\rangle_p^1|0\rangle_s^1 - i\beta\delta|0\rangle_p^1|1\rangle_s^1)|0\rangle_s^2|1_x\rangle_s^a \nonumber \\
&+& (i\beta\gamma|0\rangle_p^1|1\rangle_s^1 + \alpha\delta|1\rangle_p^1|0\rangle_s^1)|1\rangle_s^2|0_x\rangle_s^a\nonumber\\
&+&(-i\beta\gamma|0\rangle_p^1|1\rangle_s^1 + \alpha\delta|1\rangle_p^1|0\rangle_s^1)|1\rangle_s^2|1_x\rangle_s^a ]\nonumber \\
\label{redstate2}
\end{eqnarray}
using which it follows that
the possible spin measurements on Alice's second particle and the auxiliary
particle leads to the 
corresponding states with the respective probabilities, as listed in the 
TABLE. I.
\begin{table}
\begin{tabular}{|l|c|r|}
\hline
spin measurement     &   state of first               &  probability  \\
 of second particle      &  particle after              &  of spin  \\
and auxlary particle   &  spin measurement  & measurement\\
\hline
$|0\rangle_s^2 \otimes |0_x\rangle_s^a$    &  $ \alpha\gamma|1\rangle_p^1|0\rangle_s^1 + i\beta\delta|0\rangle_p^1|1\rangle_s^1$   &
$\frac{\alpha^2\gamma^2+\beta^2\delta^2}{2}$\\
\hline
$|0\rangle_s^2 \otimes |1_x\rangle_s^a$    &   $\alpha\gamma|1\rangle_p^1|0\rangle_s^1 - i\beta\delta|0\rangle_p^1|1\rangle_s^1$   &
$\frac{\alpha^2\gamma^2+\beta^2\delta^2}{2}$\\
\hline
$|1\rangle_s^2 \otimes |0_x\rangle_s^a$    &   $\alpha\delta|1\rangle_p^1|0\rangle_s^1 + i\beta\gamma|0\rangle_p^1|1\rangle_s^1$   &
$\frac{\alpha^2\delta^2+\beta^2\gamma^2}{2}$\\
\hline
$|1\rangle_s^2 \otimes |1_x\rangle_s^a$    &   $\alpha\delta|1\rangle_p^1|0\rangle_s^1 - i\beta\gamma|0\rangle_p^1|1\rangle_s^1$   &
$\frac{\alpha^2\delta^2+\beta^2\gamma^2}{2}$\\
\hline
\end{tabular}
\caption{\footnotesize}
\end{table}

Subsequently,  Alice classically communicates with Bob to tell him the 
results about 
her spin measurements (i.e., spin up or spin down for particle 1 and particle 
``a''). Now, Bob has to 
perform the remaining operations in order to recreate the
state (\ref{state}).
Bob has two particles -- one given by Alice and another particle (assumed to be
in a spin up state $|0\rangle^3_s$) which he holds initially. Depending on the
measurement results that Alice communicates to Bob, the following 
operations are performed by him:

{\it Case I:} When Alice's spin measurement on her second particle reveals a 
spin 
up ($|0\rangle^2_s$) state, and measurement on her auxiliary particle reveals 
a spin up ($|0_x\rangle_s^a$) state.

After receiving the first particle from Alice, Bob sends it through $50-50$ 
beam-splitter (BS2 in FIG. 1). The action of the beam-splitter on the
states $|0\rangle_p$ and $|1\rangle_p$ is given by
\begin{eqnarray}
|0\rangle_p \rightarrow \frac{1}{\sqrt{2}}[|a\rangle_p+i|b\rangle_p] \nonumber\\
|1\rangle_p \rightarrow \frac{1}{\sqrt{2}}[|b\rangle_p+i|a\rangle_p]
\label{bs2}
\end{eqnarray}

Now, Bob makes a CNOT operation where spin state of the third particle (held 
by him from the beginning) is the control qubit and the spin state of the first
particle (sent to him by Alice) is the target qubit. The combined state of the
two particles after the CNOT operation is given by
\begin{eqnarray}
|\psi\rangle^{13} = \frac{1}{\sqrt{2(\alpha^2\gamma^2+\beta^2\delta^2)}}[\alpha\gamma|b\rangle_{p}^{1} |0\rangle_{s}^{1} |0\rangle_{s}^{3} - \beta\delta
|b\rangle_{p}^{1} |1\rangle_{s}^{1} |1\rangle_{s}^{3} \nonumber\\
\phantom{xxxxxxxxxx} + i\alpha\gamma|a\rangle_{p}^{1} |0\rangle_{s}^{1} |0\rangle_{s}^{3} + i\beta\delta|a\rangle_{p}^{1} |1\rangle_{s}^{1} 
|1\rangle_{s}^{3}]
\label{bobstate}
\end{eqnarray}

Next, Bob measures spin of the first particle using the two sets of the 
Stern-Gerlach apparatus
(placed in both the paths $|a\rangle_p$ and  $|b\rangle_p$ along the $x$-axis).
There are four possible outcomes of Bob's measurement, i.e., 
$|a\rangle_p^1 \otimes |0_x\rangle_s^1$, $|a\rangle_p^1 \otimes |1_x\rangle_s^1$,
$|b\rangle_p^1 \otimes |0_x\rangle_s^1$, and $|b\rangle_p^1 \otimes |1_x\rangle_s^1$, all of them occurring with equal probability $1/4$. Depending on the 
exact outcome, Bob performs a corresponding unitary operation, described in
the TABLE. II, to recreate the state (\ref{state}) initially held
by Alice.
 \begin{table}
\begin{tabular}{|l|c|r|}
\cline{1-2}
\hline
Path and spin  & Unitary & final state of  \\
measurement  &operation &  Bob's particle $|\psi^{out}\rangle$\\
\hline
&     &  \\
$|a\rangle_p^1 \otimes |0_x\rangle_s^1$  & I  & $\frac{(\alpha\gamma|0\rangle_s^3+\beta\delta|1\rangle_s^3)}{\sqrt{\alpha^2\gamma^2+\beta^2\delta^2}}$\\ 
\hline
 & & \\
$|a\rangle_p^1 \otimes |1_x\rangle_s^1$& $\sigma_z$  & $\frac{(\alpha\gamma|0\rangle_s^3+\beta\delta|1\rangle_s^3)}{\sqrt{\alpha^2\gamma^2+\beta^2\delta^2}}$ \\
\hline
& &  \\
$|b\rangle_p^1 \otimes |0_x\rangle_s^1$  & $\sigma_z$  & $ \frac{(\alpha\gamma|0\rangle_s^3+\beta\delta|1\rangle_s^3)}{ \sqrt{\alpha^2\gamma^2+\beta^2\delta^2}}$\\
\hline
 &   &  \\
$|b\rangle_p^1 \otimes |1_x\rangle_s^1$  & I  &$\frac{(\alpha\gamma|0\rangle_s^3+\beta\delta|1\rangle_s^3)}{\sqrt{\alpha^2\gamma^2+\beta^2\delta^2}}$\\
\hline
\end{tabular}
\caption{\footnotesize}
\end{table}

The fidelity of the state transfer process is given by
 \begin{eqnarray}
F = |{\langle\psi^{in}|\psi^{out}\rangle}|^2 = \frac{(\alpha\gamma^2+\beta\delta^2)^2}{\alpha^2\gamma^2+\beta^2\delta^2}
\label{fid1}
\end{eqnarray} 

{\it Case II:} When Alice's spin measurement on his second particle reveals
a spin down ($|1\rangle_s^2$) state, and measurement on the auxiliary particle
 reveals a spin down ($|1_x\rangle_s^a$) state.

Bob follows same procedure similar as in Case I. He first lets the particle
sent by Alice fall on a $50-50$ beam splitter, makes a CNOT operation with
it and the particle held by him, and then measures the spin of the first
particle using his Stern-Gerlach apparatus. Finally, depending on his
measurement operation, he makes either of the four possible unitary operations
as displayed in the TABLE. III. 
\begin{table}
\begin{tabular}{|l|c|r|}
\cline{1-2}
\hline
Path and spin     &   Unitary        &    final state of  \\
measurement    &    operation   &    Bob's particle $|\psi^{out}\rangle$\\
\hline
&     &  \\
$|a\rangle_p^1 \otimes |0_x\rangle_s^1$  & $\sigma_z$  & $\frac{(\alpha\gamma|0\rangle_s^3+\beta\delta|1\rangle_s^3)}{\sqrt{\alpha^2\gamma^2+\beta^2\delta^2}}$\\
\hline
 & & \\
$|a\rangle_p^1 \otimes |1_x\rangle_s^1$& I  & $\frac{(\alpha\gamma|0\rangle_s^3+\beta\delta|1\rangle_s^3)}{\sqrt{\alpha^2\gamma^2+\beta^2\delta^2}}$ \\
\hline
& &  \\
$|b\rangle_p^1 \otimes |0_x\rangle_s^1$  & I  & $ \frac{(\alpha\gamma|0\rangle_s^3+\beta\delta|1\rangle_s^3)}{ \sqrt{\alpha^2\gamma^2+\beta^2\delta^2}}$\\
\hline
 &   &  \\
$|b\rangle_p^1 \otimes |1_x\rangle_s^1$  & $\sigma_z$  &$\frac{(\alpha\gamma|0\rangle_s^3+\beta\delta|1\rangle_s^3)}{\sqrt{\alpha^2\gamma^2+\beta^2\delta^2}}$\\
\hline
\end{tabular}
\caption{\footnotesize}
\end{table}

The fidelity of the state transfer process in this case is given by 
\begin{eqnarray}
F = |{\langle\psi^{in}|\psi^{out}\rangle}|^2 = \frac{(\beta\gamma^2+\alpha\delta^2)^2}{\beta^2\gamma^2+\alpha^2\delta^2}
\label{fid2}
\end{eqnarray}

{\it Case III:} When Alice's spin measurement on her second particle reveals a 
spin down ($|1\rangle_s^2$ ) state and measurement on the auxiliary particle
 reveals a spin up ($|0_x\rangle_s^a$) state.
\begin{table}
\begin{tabular}{|l|c|r|}
\cline{1-2}
\hline
Path and spin  &  Unitary  &  final state of\\
measurement  &  operation  &Bob's particle $|\psi^{out}\rangle$\\
\hline
  & &  \\
$|a\rangle_p^1 \otimes |0_x\rangle_s^1$  &  $\sigma_x$  & $\frac{(\beta\gamma|0\rangle_s^3+\alpha\delta|1\rangle_s^3)}{\sqrt{\beta^2\gamma^2 + \alpha^2\delta^2}}$\\
\hline
& &  \\
$|a\rangle_p^1 \otimes |1_x\rangle_s^1$  &  $\sigma_y$  & $\frac{(\beta\gamma|0\rangle_s^3 + \alpha\delta|1\rangle_s^3)}{\sqrt{\beta^2\gamma^2 + \alpha^2\delta^2}}$ \\
\hline
 &  &  \\
$|b\rangle_p^1 \otimes |0_x\rangle_s^1$  &  $\sigma_y$  & $ \frac{(\beta\gamma|0\rangle_s^3 + \alpha\delta|1\rangle_s^3)}{\sqrt{\beta^2\gamma^2 + \alpha^2\delta^2}}$ \\
\hline
  & &  \\
$|b\rangle_p^1 \otimes |1_x\rangle_s^1$  & $ \sigma_x$  & $\frac{(\beta\gamma|0\rangle_s^3 + \alpha\delta|1\rangle_s^3)}{\sqrt{\beta^2\gamma^2 + \alpha^2\delta^2}}$\\
\hline
\end{tabular}
\caption{\footnotesize}
\end{table}

Bob again follows a similar procedure as in the Cases I and II. The possible
unitary operations that he has to perform are listed in TABLE IV. The fidelity
of state transfer could also be easily obtained as in the earlier cases.

{\it Case IV:} When Alice's spin measurement on her second particle reveals a 
spin down ($|1\rangle_s^2$ ) state and measurement on the auxiliary particle 
reveals a spin down ($|1_x\rangle_s^a$) state.

This case is similar to the Case III  except in the unitary operations 
performed by Bob, which are listed in TABLE V.

\begin{table}
\begin{tabular}{|l|c|r|}
\cline{1-2}
\hline
Path and spin  &  Unitary  &  final state of\\
measurement  &  operation  &Bob's particle $|\psi^{out}\rangle$\\
\hline
  & &  \\
$|a\rangle_p^1 \otimes |0_x\rangle_s^1$  &  $\sigma_y$  & $\frac{(\beta\gamma|0\rangle_s^3+\alpha\delta|1\rangle_s^3)}{\sqrt{\beta^2\gamma^2 + \alpha^2\delta^2}}$\\
\hline
& &  \\
$|a\rangle_p^1 \otimes |1_x\rangle_s^1$  &  $\sigma_x$  & $\frac{(\beta\gamma|0\rangle_s^3 + \alpha\delta|1\rangle_s^3)}{\sqrt{\beta^2\gamma^2 + \alpha^2\delta^2}}$ \\
\hline
 &  &  \\
$|b\rangle_p^1 \otimes |0_x\rangle_s^1$  &  $\sigma_x$  & $ \frac{(\beta\gamma|0\rangle_s^3 + \alpha\delta|1\rangle_s^3)}{\sqrt{\beta^2\gamma^2 + \alpha^2\delta^2}}$ \\
\hline
  & &  \\
$|b\rangle_p^1 \otimes |1_x\rangle_s^1$  & $ \sigma_y$  & $\frac{(\beta\gamma|0\rangle_s^3 + \alpha\delta|1\rangle_s^3)}{\sqrt{\beta^2\gamma^2 + \alpha^2\delta^2}}$\\
\hline
\end{tabular}
\caption{\footnotesize}
\end{table}
Considering all the four cases together, it follows from Eqs.(\ref{fid1}) and
(\ref{fid2}), and the expressions corresponding to the cases III and IV   
that the average fidelity of state transfer is given by
\begin{eqnarray}
F_{av} = \gamma^4+\delta^4+4\alpha\beta\gamma^2\delta^2
\label{avfid}
\end{eqnarray}
If the path-spin entangled state of Alice's first particle is a maximally 
entangled state, i.e., $\alpha=\beta=\frac{1}{\sqrt{2}}$, then the average 
fidelity is equal to $1$. In this case our protocol of
state transfer is perfect, and the state of  Bob's particle's after the
completion of the protocol is the same as Alice's unknown quantum state 
which she
possesses initially.

To summarize, in this work we have shown how the information encoded
in the entanglement between
two different degrees of freedom of the same particle can be
used as a resource for performing the state transfer of an unknown qubit state
to a distant location.
This protocol is accomplished by a series of operations involving 
beam-splitters, a spin-flipper, CNOT gates, spin measurements by 
Stern-Gerlach devices, and unitary transformations. Our protocol may be
viewed as a variant of the standard teleportation scheme for a single qubit.
The 
difference here is that since the {\it intra-particle} entanglement which is
used as a resource here cannot be initially shared between the two distant
parties, the particle itself has to be transferred from Alice to Bob at 
some stage.  Note however, that the particle whose state is teleported remains
with Alice, and its initial state is destroyed by Alice's measurement, thus
avoiding any conflict with the no-cloning theorem. 
It may be noted here that the act
of physically sending one or more particles across distances is an 
unavoidable component of information theoretic protocols involved with
setting-up entanglement over distances. Whereas, in the standard teleportation 
scheme,
this process has to be initiated at the beginning in order to set-up a 
shared entangled state between two parties, in the present scheme involving
path-spin entanglement of a single particle, the particle is sent from Alice 
to Bob in the middle of
the protocol.  

Now, it is natural to ask the question as to 
what happens if the particle is
lost in transit, i.e.,  is the information about the state to be teleported
lost too ?  We show here that even if the particle is intercepted, by say,
a different
 receiver Eve (in stead of Bob) it is not possible for Eve to extract the
information encoded in the sent qubit.  Let us first re-express
Eq.(\ref{redstate}) as
\begin{eqnarray}
|\psi\rangle^{12a}&=&
 \alpha|1\rangle_p^1|0\rangle_s^1(\gamma|0\rangle_s^2 + \delta|1\rangle_s^2)|0\rangle_s^a\nonumber\\
&+& i\beta|0\rangle_p^1|1\rangle_s^1(\gamma|1\rangle_s^2 + \delta|0\rangle_s^2)|1\rangle_s^a
\label{redstate21}
\end{eqnarray}
In this scenario, the path-spin entangled qubit (particle-1) is held by Eve
and the spin qubit (particle-2) and the auxiliary particle ``a'' are 
possessed by Alice. Thereafter, Eve can
perform
measurement on the spin state of received path-spin entangled state to extract
information about the state given in Eq. (\ref{state}). But it is clear from
the above Eq. (\ref{redstate21}) that whatever be the outcome of her
measurement,
she is unable to get any information about the state given in
Eq. (\ref{state}). Note that Eve could have been be successful in her task if
Alice performs her measurement on the state of particle-2 before sending the
particle-1 towards Bob. Further, it is also possible for
Alice to retrieve the unknown state, as follows. When Bob confirms to Alice 
that he didn't get the particle, Alice makes a spin measurement on her 
auxiliary particle
in the basis \{$|0\rangle_s^a, |1\rangle_s^a$\}. According to the measurement 
outcome, she performs a suitable unitary transformation on her second particle 
(i.e., either (i) she does nothing if she gets $|0\rangle^a_s$, or (ii) she
makes the unitary operation $\sigma_x$ if she gets $|1\rangle^a_s$)  to 
retrieve the unknown state to be teleported. Note that the role of the auxiliary
particle that we have used in this protocol is to ensure that information of
the unknown state to be teleported is never lost, even if Alice's particle
1 is lost in transit.

We conclude by observing that creating the intra-particle path-spin entanglement
could be considerably easier using beam-splitters and spin-flippers, as
we have shown, than generating inter-particle entanglement through the
controlled interaction of two particles. Since one does not have to preserve
entanglement between two distant parties, our scheme should be less 
susceptible to
decoherence effects, and thus provides an advantage
over the standard scheme using two entangled qubits. The present work, however,
is limited to showing the possibility of using intra-particle entanglement
as a resource for information transfer, and issues regarding practical
feasibility need to be worked out in more details. The 
path (or linear momentum) degrees of freedom for physical particles
are always present in any experimental set-up. Here we have exploited these
path variables to first generate path-spin entanglement at the level of
a single particle, and then use it as physical resource for performing 
teleportation. This opens up the possibility of exploiting path-spin
intra-particle entanglement for performing further information theoretic
tasks. It may be also noted that though our protocol is demonstrated
here for spin-$1/2$ particles such as neutrons, it could be easily
implemented to other types of quanta such as photons using suitable optical
devices.  Finally, our analysis  generally reemphasizes the notion 
that entanglement is a fundamental concept independent of either any particular
physical realization of Hilbert space \cite{hybrid}, or delocalisability of
the involved modes, and specifically highlights that hybrid
entanglement at the level of a single particle \cite{home,hasegawa}
could be regarded as a real
physical resource.
  
{\bf  Acknowledgements} ASM and DH acknowledge support from the DST project
no. SR/S2/PU-16/2007. DH thanks the Centre for Science and Consciousness, 
Kolkata, India.

\end{document}